\documentclass[%
reprint,
superscriptaddress,
 amsmath,amssymb,
 aps,
]{revtex4-1}

\usepackage{graphicx,setspace}
\usepackage{dcolumn}
\usepackage{bm}
\usepackage[mathlines]{lineno}
\usepackage[colorlinks=true,citecolor=blue,linkcolor=red,anchorcolor=green,urlcolor=cyan]{hyperref}
\usepackage{mathrsfs}
\usepackage{xcolor}
\usepackage{float}
\usepackage{ulem}
\usepackage[utf8]{inputenc}

\newcommand{\be}{\begin{equation}}
\newcommand{\ee}{\end{equation}}
\newcommand{\bal}{\begin{aligned}}
\newcommand{\eal}{\end{aligned}}

\newcommand{\bea}{\setlength\arraycolsep{2pt} \begin{eqnarray}}
\newcommand{\eea}{\end{eqnarray}}

\graphicspath{{Bilder/}}

\begin{document}


\title{Shadows of the accelerating black holes}

\author{Ming Zhang}
\email{mingzhang@jxnu.edu.cn}
\affiliation{Department of Physics, Jiangxi Normal University, Nanchang 330022, China}
\author{Jie Jiang}
\email{jiejiang@mail.bnu.edu.cn (corresponding author)}
\affiliation{Department of Physics, Beijing Normal University, Beijing 100875, China}

\date{\today}

\begin{abstract}
Due to the acceleration of the black hole, the circular orbits of the photons will deviate from the equatorial plane and the property of the black hole shadow will change. We find that the latitude of the circular orbit increases with the increasing acceleration and then show that the observer's inclination angles which make the shadow radius and the shadow distortion maximum increase with the increasing acceleration for the accelerating Kerr black hole.

\end{abstract}


\maketitle


\section{Introduction}

Recently, the image of the supermassive black hole located at the centre of M87 galaxy was taken by the Event Horizon Telescope (EHT) Collaboration \cite{Akiyama:2019cqa,Akiyama:2019brx,Akiyama:2019sww,Akiyama:2019bqs,Akiyama:2019fyp,Akiyama:2019eap}. The prediction of the general relativity was supported by the observation result if the image was modelled by Kerr geometry. However, due to the finite resolution of the image, there still exist possibilities of modelling it by other geometries. Nowadays we cannot verdict which gravity theory is closer to observation, however, with the developments of observation, such as the BlackHoleCam \cite{Goddi:2017pfy}, it will be possible.

The  null-like photons emitted by sources everywhere except the region between the observer and the black hole  flying past a black hole have three kinds of destinies, which are absorbing by the black hole, reflecting by the black hole and a state between them, which means that the photon will revolve around the black hole. The black hole shadow is defined as the dark sky of an observer positioning at spatial finite or infinite distance.

Early in the 1960s, the shadow radius of the spherically symmetric Schwarzschild black hole as a black circular disk was calculated in \cite{synge1966escape}, and later \cite{luminet1979image} provided a formula calculating the size of the shadow of the Schwarzschild black hole surrounded by an accretion disk. For a rotating black hole, the shadow is elongated silhouette-like  in the direction of the rotating axis, which results from the dragging effect of the rotation. The shadow of a Kerr black hole was studied in \cite{hawking1973black}. Shadows of numerous kinds of spacetime geometries have been investigated since then, including black hole ones \cite{Hioki:2008zw,Grenzebach:2014fha,Wang:2017qhh,Guo:2018kis,Li:2020drn,Yan:2019etp,Hennigar:2018hza,Konoplya:2019sns,Bambi:2008jg,Bambi:2010hf,Konoplya:2019fpy,Wei:2019pjf,Wei:2018xks,Wang:2018prk,Liu:2020ola,Kumar:2019pjp}, and wormhole ones \cite{Ohgami:2015nra,Nedkova:2013msa,Shaikh:2018kfv,Amir:2018szm,Amir:2018pcu,Wang:2020emr,Chang:2020lmg,Gralla:2020yvo}. For a review, see \cite{Cunha:2018acu}. It has been shown that the apparent characteristics (the size, the distortion) of the black hole shadows vary with the parameters of the black holes.

Thus far, the spacetime geometries that have been investigated share similar characteristics. The circular orbits of the photons around the rotating black holes in those geometries are on the equatorial plane. The sizes and the distortions of the shadows for those black holes are maximum if the inclination angles of the observers are $\pi/2$. However, in the black hole solution family of the Einstein's general relativity, there is a special kind of spacetime geometry, dubbed as C-metric \cite{Weyl:1917gp,Kinnersley:1970zw}, which belongs to  the well-known Plebański-Demiański spacetime \cite{Griffiths:2005qp}. The rotating C-metric \cite{Plebanski:1976gy} describes the accelerating Kerr black hole. Due to the acceleration of the black hole, the circular orbits of the photons as well as the properties of the shadow will change in unique ways. 

We in this paper will introduce our investigations of the shadows for the accelerating Kerr black holes. In Sec. \ref{sec2}, we will investigate the characteristics of the circular orbits of the photons around the accelerating Kerr black holes.  We will show the contour of the accelerating Kerr black hole observed by a zero-angular-momentum-observer at a finite distance in Sec. \ref{sec3} and analyze the properties of the observables for the shadows in Sec. \ref{sec4}. The last section will be devoted to our closing remarks. 

It is worth mentioning that the shadows of the accelerating black holes for a Carter observer were previously studied in \cite{Grenzebach:2015oea}. But our study in this paper emphasizes the special effects of the black hole acceleration that was not discussed previously and we suspect that the results are also suitable for the shadows obtained in the Carter frame.

\begin{figure*}[htpb!]
\begin{center}
\includegraphics[width=1.6in,angle=0]{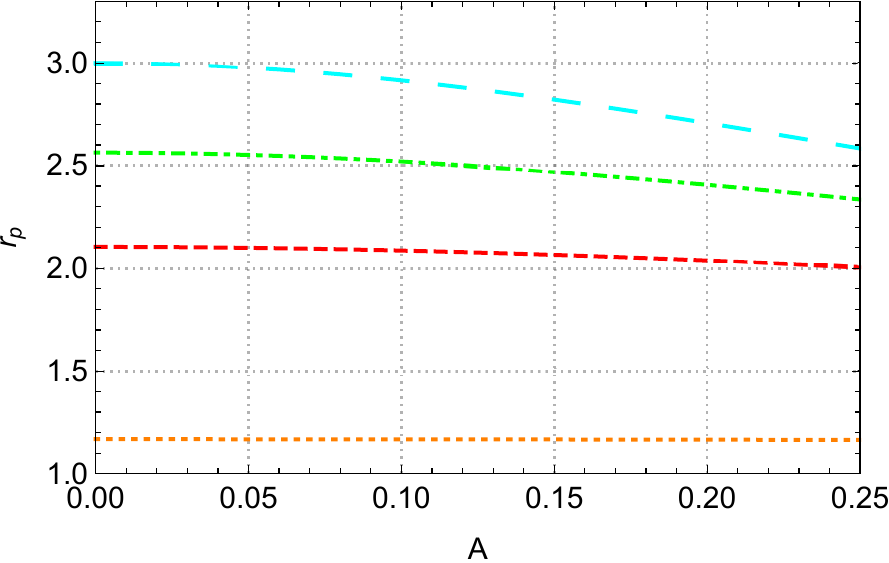}
\includegraphics[width=1.6in,angle=0]{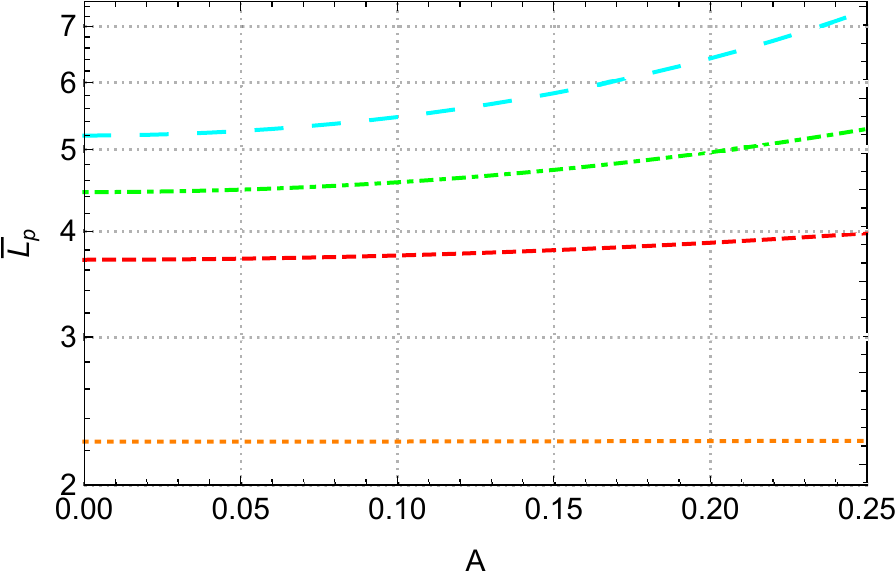}
\includegraphics[width=1.6in,angle=0]{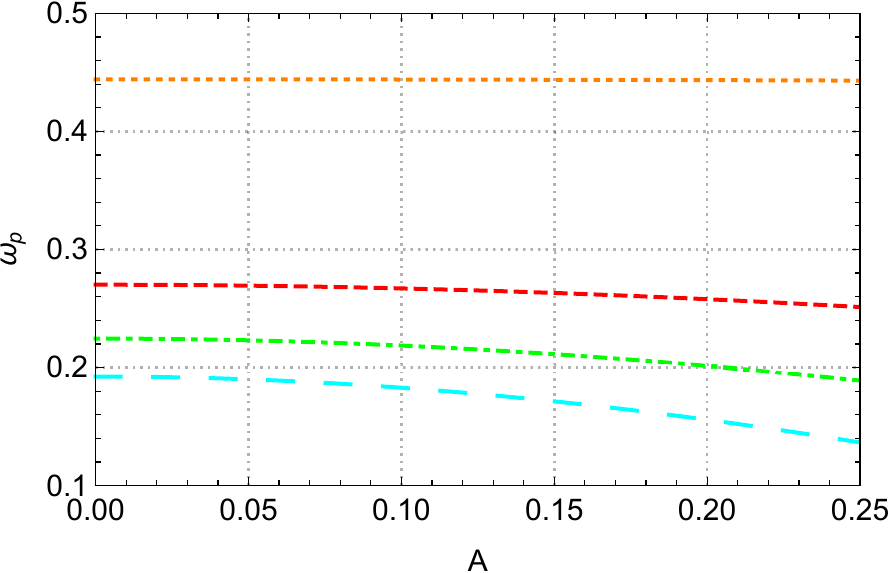}
\includegraphics[width=2.1in,angle=0]{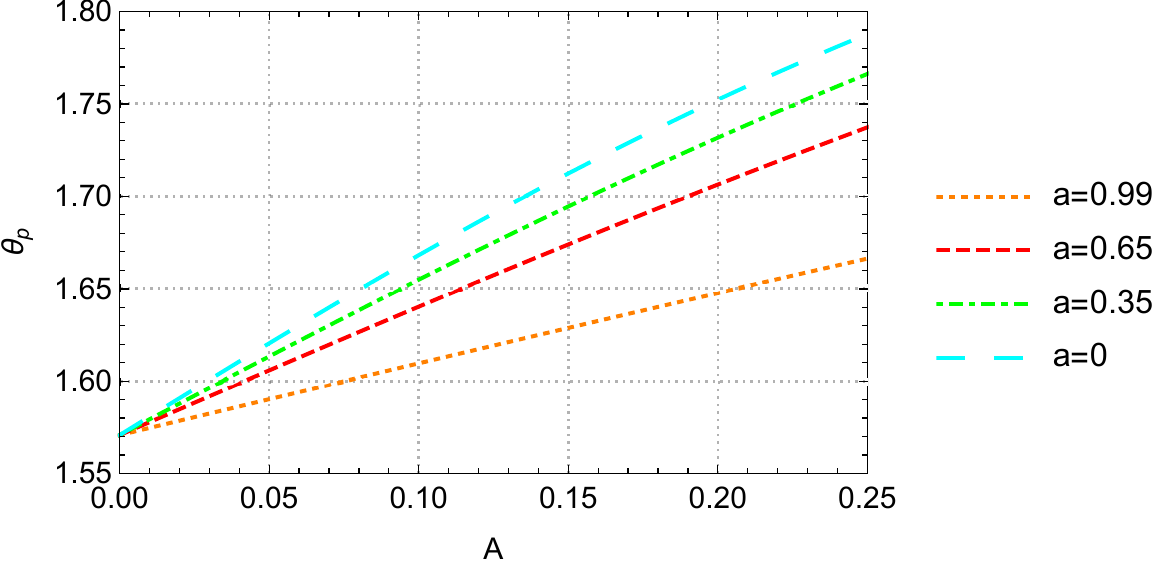}\\
\includegraphics[width=1.6in,angle=0]{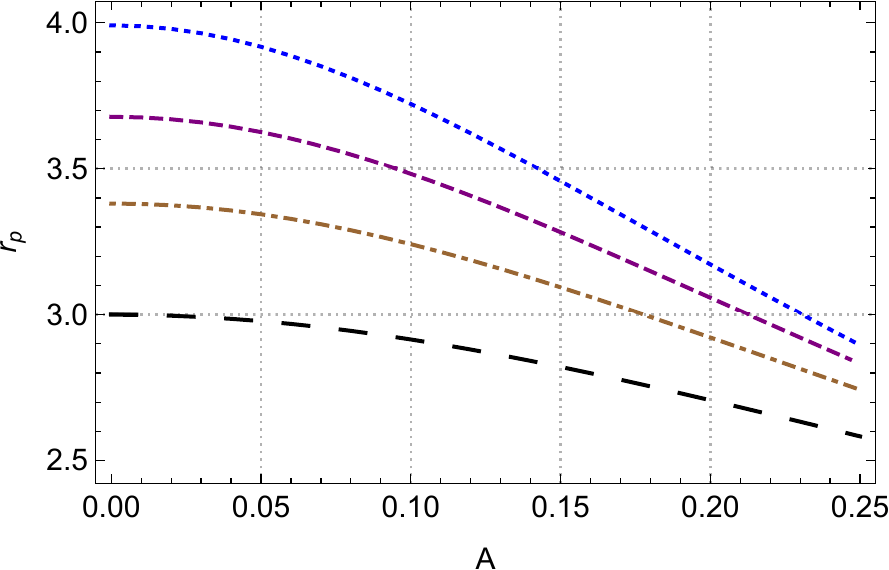}
\includegraphics[width=1.6in,angle=0]{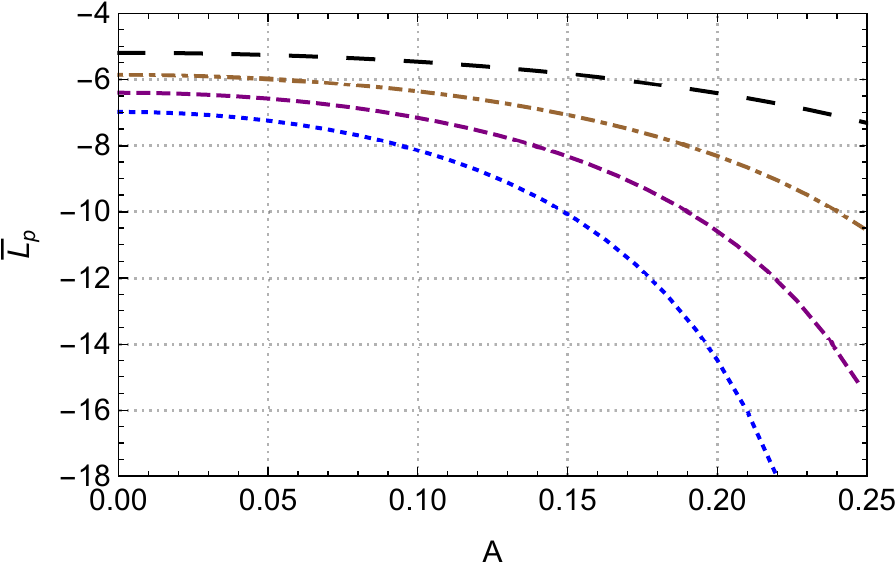}
\includegraphics[width=1.6in,angle=0]{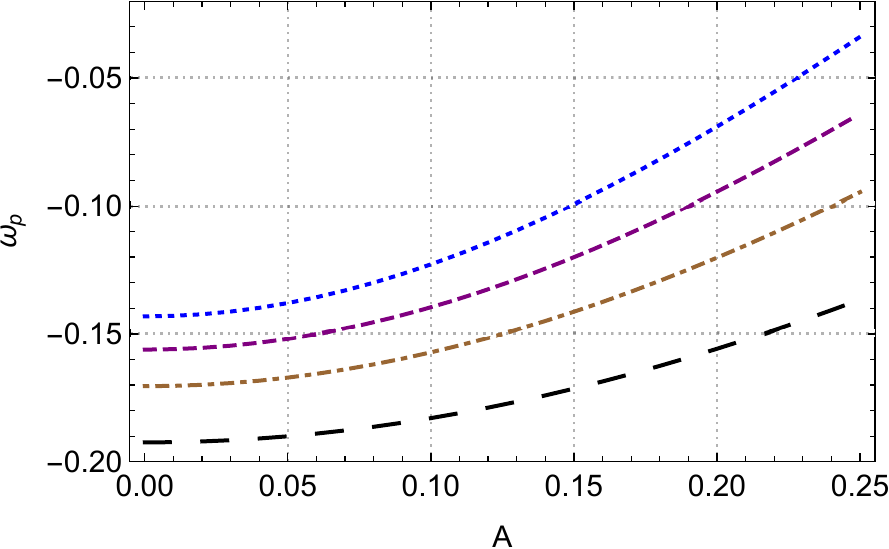}
\includegraphics[width=2.1in,angle=0]{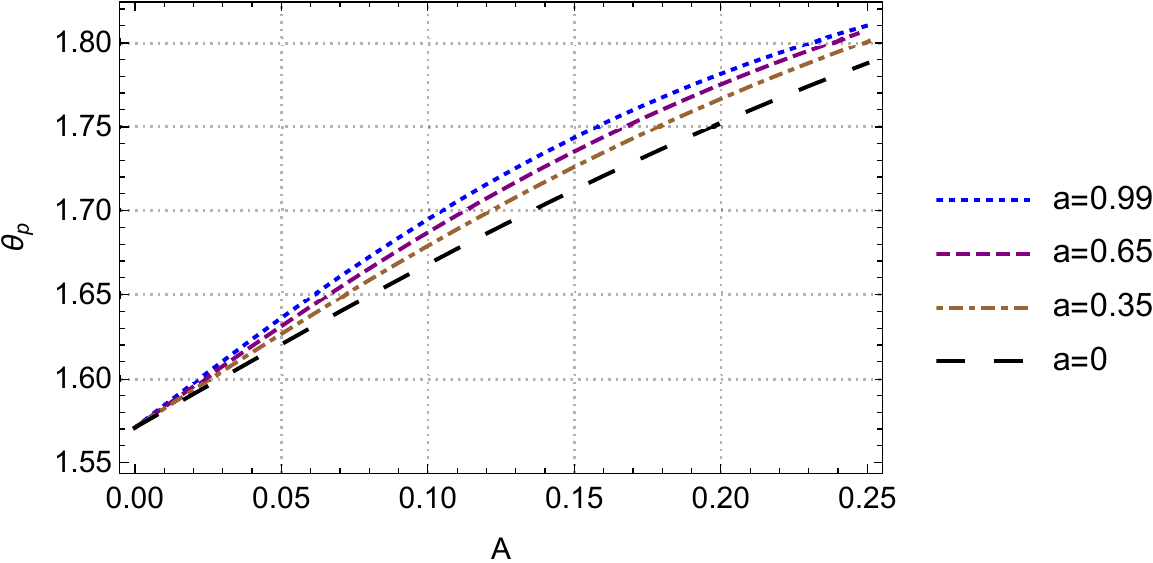}
\end{center}
\vspace{-5mm}
 \caption {Variations of the circular orbit parameters for the photon around the accelerating Kerr black hole with $m=1\,,E=1.$. The upper diagrams are for the prograde orbits and the bottom ones are for the retrograde orbits.}\label{pp}
\end{figure*}

\section{Circular orbits for the photons around the accelerating Kerr black hole}\label{sec2}
The line element of the accelerating Kerr black hole is \cite{Griffiths:2005qp}
\be\label{metric}
\bal
ds^2=&\frac{1}{\Omega^2}\left[\Sigma  \left(\frac{d\theta^2}{\Delta _{\theta }}+\frac{dr^2}{\Delta _r}\right)-\frac{ \left(\Delta _r-a^2 \Delta _{\theta } \sin ^2\theta \right)dt^2}{\Sigma }\right.\\ &\left.+\frac{2  \left[\chi  \Delta _r-a \Delta _{\theta } \sin ^2\theta (a \chi +\Sigma )\right] dt d\phi}{\Sigma }\right.\\&\left.+\frac{ \left[\Delta _{\theta } \sin ^2\theta  (a \chi +\Sigma )^2-\chi ^2 \Delta _r\right] d\phi^2}{\Sigma }\right],
\eal
\ee
where
\be\bal
\chi&=a\sin ^2\theta,\\
\Omega&=1-A r \cos\theta,\\
\Sigma&=r^{2}+a^2 \cos ^2\theta ,\\
\Delta_r&=(1-A^2r^2)(r^2-2m r+a^2),\\
\Delta_\theta&=1-2 A m \cos \theta+ a^2  A^2  \cos ^2\theta .\nonumber
\eal\ee
$m,\,a$ are mass parameter, rotation parameter of the black hole. $A$ represents the acceleration of the black hole. $\Omega>0$ is the conformal factor and it becomes zero at the conformal boundary $r_A=1/|A\cos\theta|$. The event horizon $r_{+}$ of the black hole makes $\Delta_{r}(r_{+})=0$, which should theoretically satisfy $r_+<r_A$.

The null-like geodesic equation for the photon in the accelerating Kerr black hole spacetime reads \cite{Grenzebach:2015oea}
\begin{equation}
\frac{\Sigma}{\Omega^{2}}\frac{dt}{d\tau}= \frac{\chi(L_{z}-E\chi)}{\Delta_{\theta} \sin^{2}\theta}
+\frac{(\Sigma + a\chi) \left[\Sigma + a\chi)E - aL_{z}\right]}{\Delta_{r}},
\end{equation}
\begin{equation}
\frac{\Sigma}{\Omega^{2}}\frac{d\phi}{d\tau}= \frac{L_{z}-E\chi}{\Delta_{\theta} \sin^{2}\theta}
		+\frac{a\left[\Sigma + a\chi)E - aL_{z}\right]}{\Delta_{r}},
\end{equation}
\begin{equation}
\left(\frac{\Sigma}{\Omega^{2}}\right)^{2}\left( \frac{d\theta}{d\tau}\right)^2 = \Delta_{\theta}K - \frac{(\chi E - L_{z})^{2}}{\sin^{2}\theta} \equiv \Theta(\theta),\label{loeff}
\end{equation}
\begin{equation}
\left(\frac{\Sigma}{\Omega^{2}}\right)^{2}\left( \frac{dr}{d\tau}\right)^2 =\left[(\Sigma + a\chi)E-aL_{z}\right]^{2} - \Delta_{r}K\equiv R(r),\label{reff}
\end{equation}
where $\tau$ is the affine parameter, $E$ and $L_z$ are the energy and the angular momentum corresponding to the Killing vectors $\partial_t$ and $\partial_\varphi$, respectively. $K$ is the Carter constant \cite{Carter:1968rr}. We have defined the longitudinal effective potential as $\Theta(\theta)$ and the radial effective potential as $R(r)$.

Generally, the physically reasonable region for the motion of the photons are restricted by $R(r)\geqslant 0$ and $\Theta (\theta)\geqslant 0$. If additional conditions $R'(r)=0$ and $\dot{\Theta}(\theta)=0$ are proposed, where the $'$ and ${}^{\cdot}$ indivisually denote the derivative with respect to coordinates $r$ and $\theta$, the motion of the photon is confined to a circular orbit with a constant latitudinal angle. We will study the properties of the parameters for this circular orbit. In other words, denoting the radius and the latitude as $r_p$ and $\theta_p$, the circular orbit of the photon around the accelerating Kerr black hole in the domain of outer communication ($\Delta_r>0$) complies with the conditions
\begin{equation}
R(r_p)=0,\quad R'(r_p)=0,\quad  R''(r_p)>0,\label{co1}
\end{equation}
\begin{equation}
\Theta(\theta_p)=0,\quad \dot{\Theta}(\theta_p)=0,\quad \ddot\Theta(\theta_p)<0,\label{co2}
\end{equation}
where $R''(r_p)>0$ indicates that the circular orbit in the domain of outer communication is radically unstable (to see this, we can refer to, e.g., \cite{chandrasekhar1985mathematical} for the non-accelerating case). Besides, in \cite{Grenzebach:2015oea}, stable spherical light-rays are shown in the region with $\Delta_r<0$ for the accelerating Kerr black holes, which is not our interests in this paper. 

We can obtain
\begin{equation}
\bar{K} = \frac{16r^{2}\Delta_{r}}{(\Delta_{r}')^{2}}=\frac{1}{a}\left( \Sigma + a\chi - \frac{4r\Delta_{r}}{\Delta_{r}’}\right)
\end{equation}
and
\begin{equation}
\bar{L} =\frac{1}{a}\left( \Sigma + a\chi - \frac{4r\Delta_{r}}{\Delta_{r}’}\right)
\end{equation}
from Eq. (\ref{co1}). The Eq. (\ref{co2}) gives
\begin{equation}
\bar{K} = \frac{(\chi  - L_{E})^{2}}{\Delta_\theta\sin^{2}\theta} =\frac{4a (\chi-L_E)\cot\theta}{\Delta_\theta^{'}},
\end{equation}
\begin{equation}
\bar{L} = \frac{\chi \Delta_\theta^{'}-2a \Delta_\theta\sin(2\theta)}{\Delta_\theta^{'}},
\end{equation}
where we have denoted
\begin{equation}
\bar{K} =\frac{K}{E^2},\quad \bar{L}=\frac{L}{E}.
\end{equation}

Due to the existence of the acceleration of the black hole, the photon cannot have a circular motion on the equatorial plane of the black hole. In this case, the circular motion of the photon is confined to a $\theta_p=\theta_c\neq \pi/2$ plane, with $\theta_c$ a constant. To see this, we just need to substitute $\theta_p=\pi/2$ into $\dot{\Theta}$. We will obtain $\dot{\Theta}(\theta_p)=2 A K m$, which cannot be vanishing unless $A=0$. so the photon on the equator will be latitudinally unstable for the accelerating black hole. Intuitively, we cannot tell whether $0<\theta_c <\pi/2$ or $\pi/2<\theta_c <\pi$ directly, so further calculation is necessary.

Fortunately, for the $a=0$ case, we can analytically solve the Eqs. (\ref{co1}) and (\ref{co2}), and obtain the circular orbit parameters as
\begin{equation}
\bar{K}_p =\frac{\left(\sqrt{12 A^2+1}-1\right)^3}{2 A^2 \left(-4 A^2+\sqrt{12 A^2+1}-1\right)^2},
\end{equation}
\begin{equation}
r_p= \frac{\sqrt{12 A^2+1}-1}{2 A^2},\label{sc1}
\end{equation}
\begin{equation}
\cos\theta_p=\frac{1-\sqrt{12 A^2+1}}{6 A},\label{sc2}
\end{equation}
and
\begin{equation}
\bar{L}_p=\pm  \sqrt{\bar{K}_p (1-2 A \cos \theta _p)}  \sin \theta _p\label{sc3},
\end{equation}
where $\bar{K}_p$, $r_p$, $\bar{L}_p$ are reduced Carter constant, radius and reduced angular momentum of the photon on the circular orbit, respectively. $``+"$ is for the prograde orbit and  $``-"$ is for the retroprograde orbit.

For the $a\neq 0$ case, we can resort to numerical calculations. In Fig. \ref{pp}, we have shown the variations of the circular orbit parameters of the photons with respect to the acceleration of the black hole, where $\omega_p$ is the angular velocity of the photon on the circular orbit, defined by $\dot{r}/\dot{\phi}$. We can check that the Eqs. (\ref{sc1}), (\ref{sc2}) and (\ref{sc3}) are in consistent with the $a=0$ case in Fig. \ref{pp}, which is obtained numerically. For the prograde orbits, we can see that the radius and the angular velocity decreases with the increasing acceleration while the angular momentum increases with the increasing acceleration. For the retrograde orbits, the variation trend of the radius is similar to the ones for the prograde orbits, but the angular momentum and angular velocity are contrary to their prograde orbits counterparts. What we concentrate on is the effect of the black hole acceleration on the angle $\theta_p$. It is obvious that $\theta_p$ increases with the increasing acceleration both for the prograde orbits and for the retrograde orbits. For a black hole with certain angular momentum and acceleration, we can know that the prograde circular orbit and the retrograde circular orbit occupy different angles $\theta_p$. It is the acceleration of the black hole that makes them separate. At the same time, we can see that the greater the black hole's acceleration is, the greater the separation is.

\begin{figure}[htbp!]
\begin{center}
\includegraphics[width=70mm,angle=0]{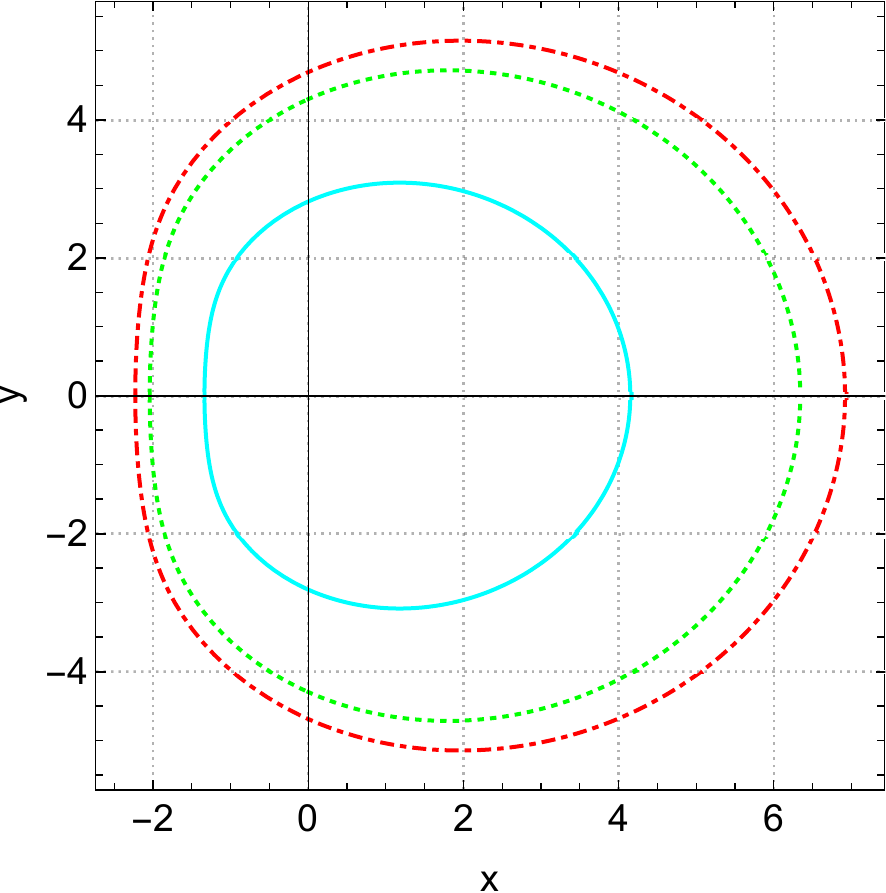}
\end{center}
\vspace{-5mm}
 \caption {The  curves are the edges of the shadows of the accelerating Kerr black holes with  $r_{O}=100,\,m=1\,,a=0.99\,,\theta_O=\pi/2$. From outside to inside we set $A=0, 0.004,\,0.008$.}\label{f1}
\end{figure}

\begin{figure}[htbp!]
\begin{center}
\includegraphics[width=70mm,angle=0]{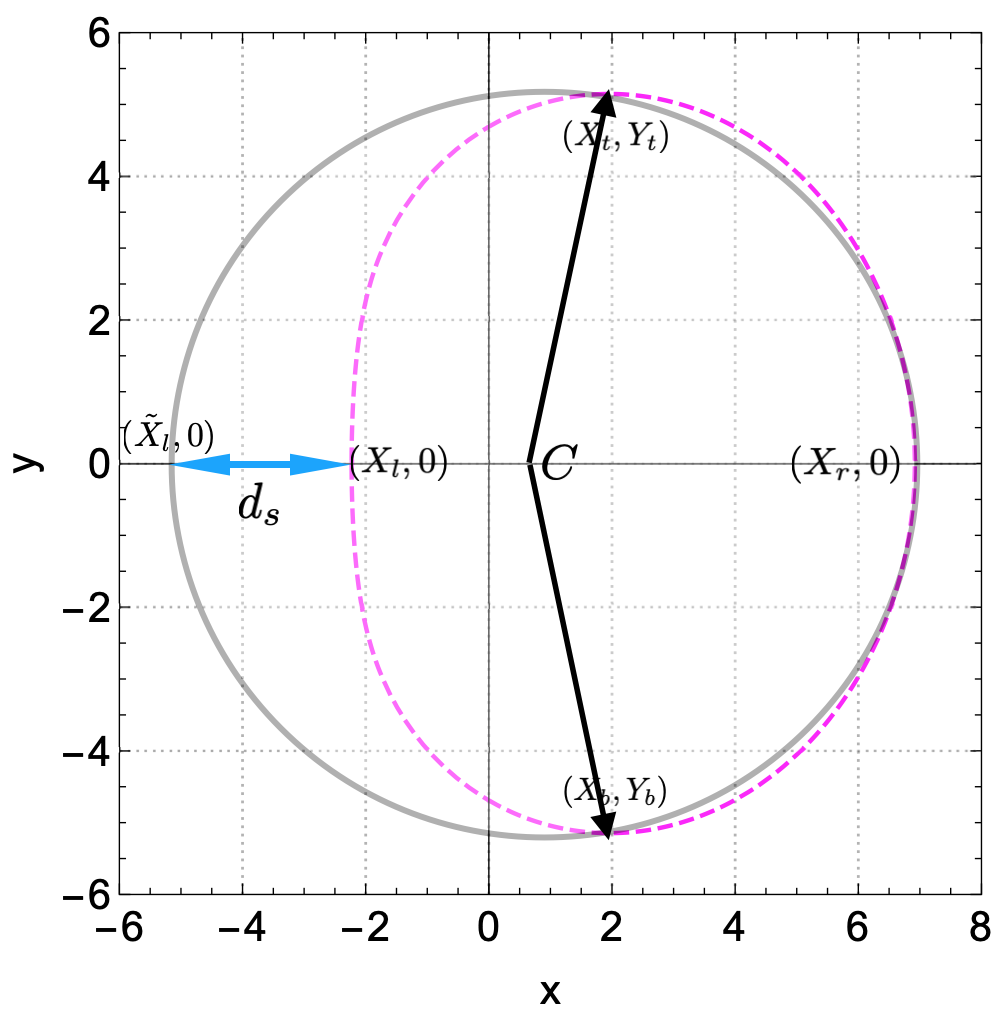}
\end{center}
\vspace{-5mm}
 \caption {A schematic picture of the rotating black hole shadow. In the picture, the rightmost, leftmost, top and bottom points on the boundary of the black hole shadow are $(X_{r}, 0),\, (\tilde{X}_{l}, 0),\,(X_{t}, Y_{t}),\,(X_{b}, Y_{b})$, respectively, where $X_t =X_b$.  The absolute values of $X_{r}$ and $\tilde{X}_{l}$ are equal if the rotation of the black hole vanishes.}\label{f0}
\end{figure}

\section{Contours of the shadows for the accelerating Kerr black holes}\label{sec3}

Now we calculate the contour of the  black hole shadow, which is also the critical curve that can be seen by an observer. We can choose the following normalized and orthogonal tetrad
\begin{subequations}
\begin{eqnarray}
  \hat{e}_{(t)} & = & \sqrt{\frac{g_{\phi \phi}}{g_{t \phi}^2 - g_{t t} g_{\phi
  \phi}}} \left( \partial_t - \frac{g_{t \phi}}{g_{\phi \phi}} \partial_{\phi}
  \right),\label{et}\\
  \hat{e}_{(r)} & = & \frac{1}{\sqrt{g_{r r}}} \partial_r,\\
  \hat{e}_{(\theta)} & = & \frac{1}{\sqrt{g_{\theta \theta}}} \partial_{\theta},\\
  \hat{e}_{(\phi)} & = & \frac{1}{\sqrt{g_{\phi \phi}}} \partial_{\phi},
\end{eqnarray}
\end{subequations}
for the static observer located at finite distance. These bases are spacelike in the domain of outer communication except the vector $\hat{e}_{(t)}$ which is timelike. This tetrad frame is named as the zero-angular-momentum-observer (ZAMO) reference frame, as one can see that $\hat{e}_{(t)}\cdot\partial_{\varphi}=0$.

Using the ZAMO tetrad, the four-momentum can be projected as 
\begin{subequations}
\begin{eqnarray}
  p^{(t)} & = & - p_{\mu} \hat{e}_{(t)}^{\mu}, \\
  p^{(i)} & = & p_{\mu} \hat{e}_{(i)}^{\mu}, \quad i = r, \theta, \phi,
\end{eqnarray}
\end{subequations}
where $i=r,\,\theta,\,\phi$. This is the four-momentum measured by the locally static observer.

\begin{figure}[htbp!]
\begin{center}
~\includegraphics[width=70mm,angle=0]{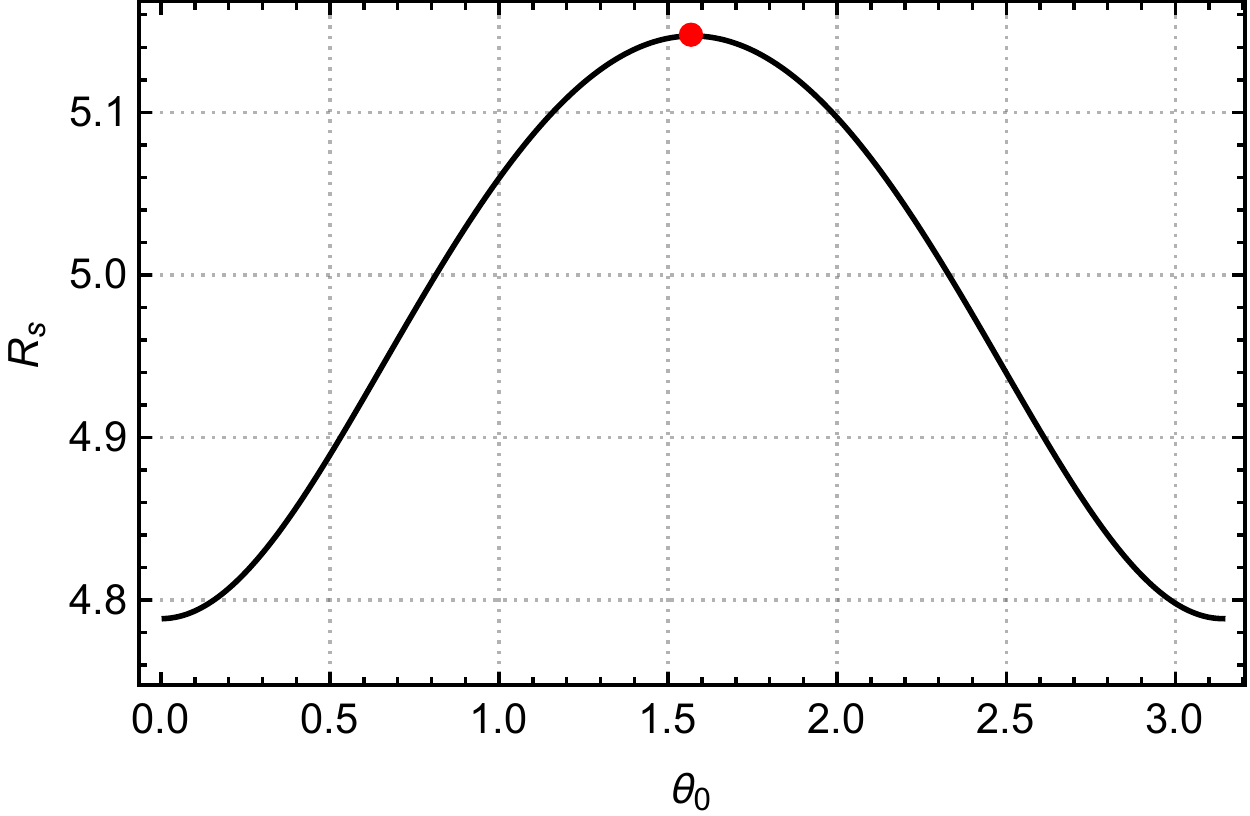}
\includegraphics[width=71.5mm,angle=0]{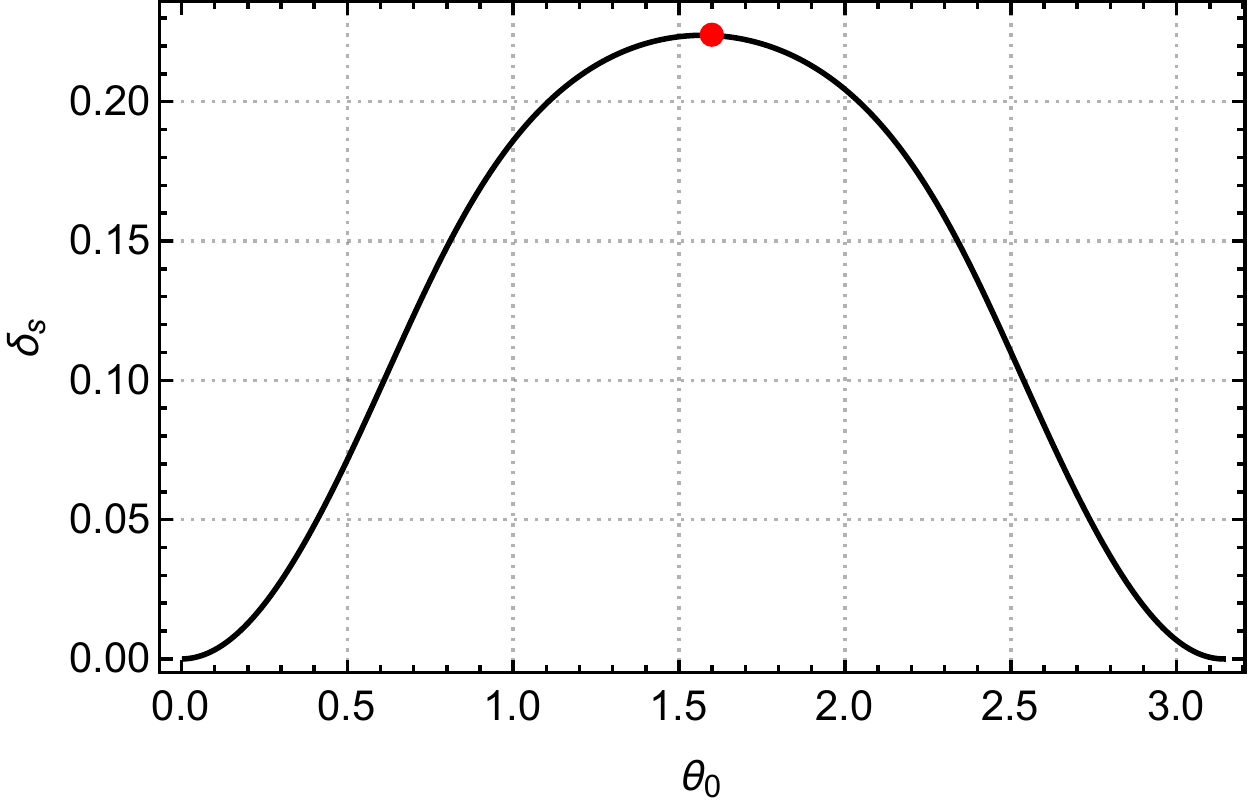}
\end{center}
\vspace{-5mm}
 \caption {Variations of the size and the distortion of the black hole shadow with respect to the inclination angles of the observer. The red points in the diagrams are the biggest size or distortion. }\label{fx}
\end{figure}

\begin{figure*}[htbp!]
\begin{center}
\includegraphics[width=70mm,angle=0]{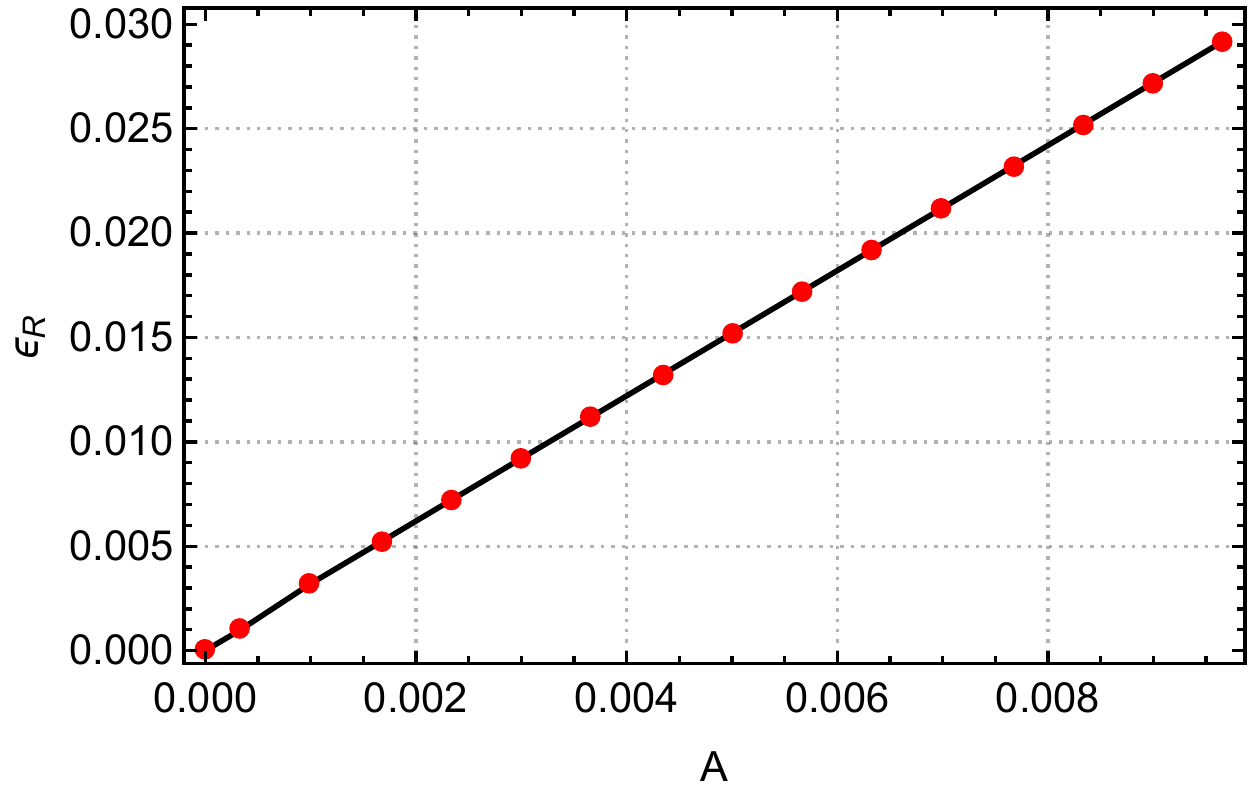}
\includegraphics[width=70mm,angle=0]{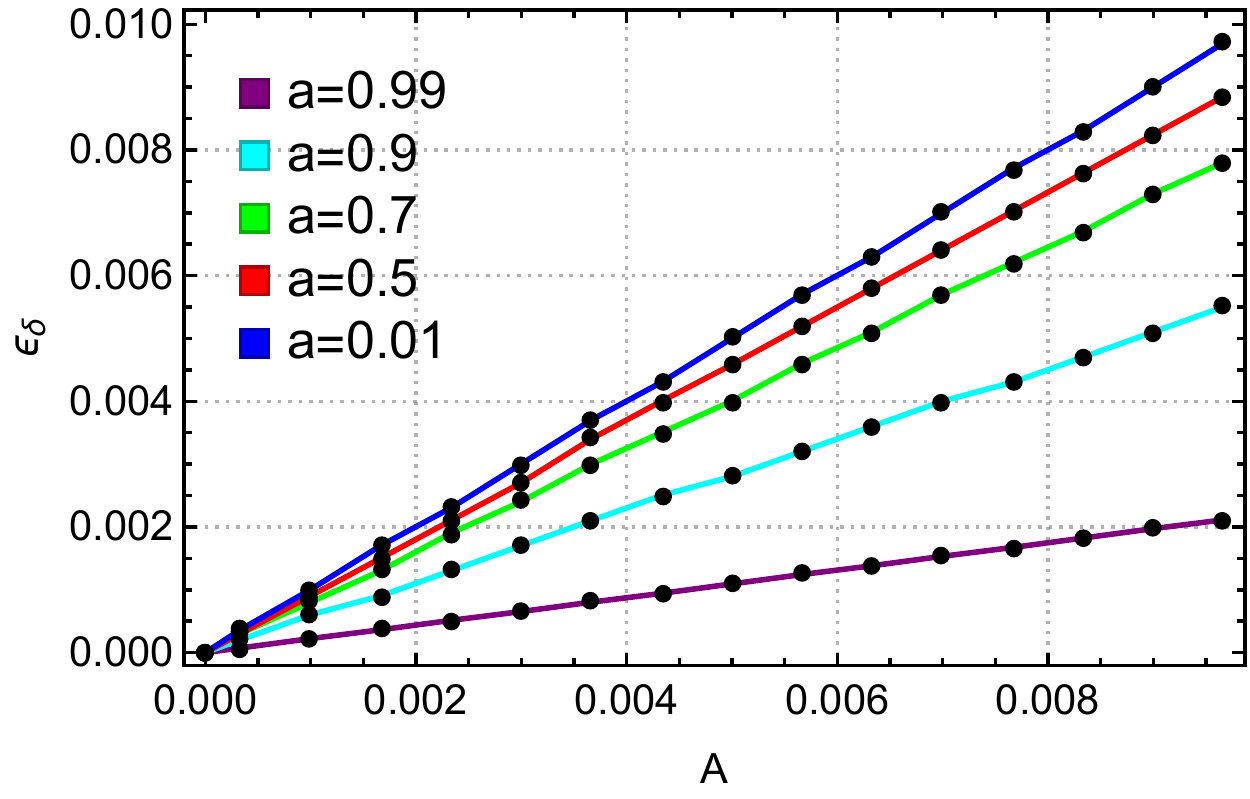}
\end{center}
\vspace{-5mm}
 \caption {The left diagram shows the variation of the degree of deviation of the inclination angle that makes the shadow radius $R_s$ maximum with respect to the acceleration of the black hole with $a=0.99$.  The right diagram shows the variation of the degree of deviation of the inclination angle that makes the shadow distortion $\delta_s$ maximum with respect to the acceleration of the black hole. We have set $m=1,\,r_+\ll r_O=100<1/A$.}
 \label{f2}
\end{figure*}

As the mass of the photon is zero, the linear momentum $\vec{P}$ as a 3-vector relates with $p^{(i)}$ just satisfies    $| \vec{P} | = p^{(t)}$ in the observer's frame and  the observation angles $(\alpha,
\beta)$ can be introduced as \cite{Cunha:2016bpi}
\begin{eqnarray}
  p^{(r)} & = & | \vec{P} | \cos \alpha \cos \beta, \\
  p^{(\theta)} & = & | \vec{P} | \sin \alpha, \\
  p^{(\phi)} & = & | \vec{P} | \cos \alpha \sin \beta.
\end{eqnarray}
Then we have
\be
\bal
\sin\alpha&=\frac{p^{(\theta)}}{p^{(t)}}\\&=\left.\pm\frac{\Omega}{\zeta-L_E \gamma}
  \sqrt{\frac{\Delta_\theta K_E\sin^2\theta-(\chi-L_E)^2}{\Sigma \Delta_\theta\sin^2\theta}}\right|_{(r_O,\theta_O)}
\eal
\ee
\be
\bal
& \tan \beta = \frac{p^{(\phi)}}{p^{(r)}}\\&=\left.\frac{L_E \sqrt{\Sigma \Delta_r}}
    {\Omega\sqrt{g_{\phi\phi}}\sqrt{((\Sigma+a^2\sin^2\theta)-aL_E)^2-\Delta_rK_E}}\right|_{(r_O,\theta_O)},
\eal
\ee
where $\zeta\equiv \hat{e}_{(t)}^t$ , $\gamma\equiv\hat{e}_{(t)}^\phi$. $r_{O}$ and $\theta_{O}$ stand for the radial position and the inclination angle between the direction of the rotation axis of the accelerating Kerr black hole and the static observer. One should note that here $\zeta$ and $\gamma$ are valued by the photon radius $r_{p}$. The Cartesian coordinate $(x, y)$ can be introduced for the apparent position on the plane of the sky for the observer, as
\begin{equation}
  x \equiv - r_O \beta, \quad y \equiv r_O \alpha .\label{xy}
\end{equation}

We show the shadow of the accelerating Kerr black hole in Figs. \ref{f1}. From the figure, we can see that the size of the black hole shadow decreases with the increasing acceleration of the black hole.

\section{Analysis of the observables}\label{sec4}

To know the black hole parameters by using observed data, it is important to study the observables. Two important quantities of the observables are the size and the distortion of the black hole. The size of the black hole can be reflected by the radius $R_{s}$ of a reference circle related with the black hole shadow and $\delta_{s}$ can be used to  measure the distortion of the black hole shadow compared with the reference circle \cite{Hioki:2009na}. We plot a schematic picture of the accelerating Kerr black hole shadow with its reference circle in Fig. \ref{f0}.  Making use of the quantities in the picture, the radius $R_s$ and the distortion $\delta_s$ as observables of the black hole can be defined as \cite{Hioki:2009na}
\begin{equation}
R_{s}=\frac{(X_{t}-X_{r})^{2}+Y_{t}^{2}}{2(X_{r}-X_{t})},
\end{equation}
\begin{equation}
\delta_{s}=\frac{d_{s}}{R_{s}}=\frac{X_{l}-\tilde{X}_{l}}{R_{s}}.
\end{equation}

The observables rely on the position of the observer. The inclination angle of the observer affects the value of the observables. For the Kerr black hole without acceleration, it is not difficult to imagine that the radius $R_s$ and the distortion $\delta_s$ get their biggest value if the inclination angle of the observer is $\theta_O=\pi/2$ in a condition of constant radial position. However, if we take the acceleration of the Kerr black hole into consideration, things do change. The acceleration of the black hole deviates the inclination angle that makes the observables maximum. As shown in the schematic picture Fig. \ref{fx}, the red points that make $\delta_s$ and $R_s$ maximum will deviates from $\theta=\pi/2$ if the black hole is accelerating. We define
\begin{equation}
\epsilon_R\equiv\theta_m (A\neq 0, R_s=R_{\text{max}})-\frac{\pi}{2}
\end{equation}
and
\begin{equation}
\epsilon_\delta\equiv\theta_m (A\neq 0, \delta_s=\delta_{\text{max}})-\frac{\pi}{2}
\end{equation}
to reflect the extent of deviation of the inclination angle that makes the observables maximum due to the acceleration of the black hole, where $\theta_m$ is the inclination angle of the observer where the observables are maximum.  First, we find that both $\epsilon_R$ and $\epsilon_\delta$ are positive, i.e., the acceleration of the black hole makes the critical inclination angle deviate from the equatorial plane to the one nearer to the south pole ($\theta=\pi$) of the black hole. Then we  quantitatively  show how the acceleration of the black hole changes the degree of deviation. It is explicitly shown in Fig. \ref{f2} that both the degree of the deviation of the shadow radius and the deviation of the shadow distortion grow with the increasing acceleration of the black hole. In the left diagram of Fig. \ref{f2}, we do not show the curves for the case of $a<0.99$, as the difference between $a=0.99$ case and $a<0.99$ case is small and we cannot differentiate them on the diagram. Nevertheless, just like the cases in the right diagram, the curve with greater $a$ is over the one with smaller $a$, i.e., the bigger the angular momentum of the black hole is, the smaller the degree of deviation is.

\section{Closing Remarks}\label{sec5}
In this paper, we studied the effects of the acceleration on the circular orbits of the photons around the accelerating Kerr black hole and the shadow observables for the accelerating Kerr black hole. We found that the latitude of the photon's circular orbit increases with the acceleration of the black hole. We also discovered that due to the appearance of the acceleration, the critical inclination angles of the observer making the size and the distortion of the shadow maximum deviate from the equator of the black hole and incline to the south pole. Quantitatively, that critical inclination angle increases with the increasing acceleration.

We now analyze the conical deficit of the accelerating Kerr black hole and try to explain the deviation of the photon's circular orbit from the equatorial plane and the deviation of the critical inclination observe angles. The conical singularities of the black hole locate at $\theta=0, \pi$. The conical singularity of the accelerating black hole originates from the existence of the deficit angle at the pole. We can assume $\phi\in [0, 2\pi C)$ in (\ref{metric}) \cite{Destounis:2020pjk}, with $C$ being a rescale factor. Then the deficit angles $\varpi_+$  at the north pole $\theta=0$ and $\varpi_-$ at the south pole $\theta=\pi$ can be respectively obtained as
\begin{equation}
\varpi_\pm=2\pi [1-C \Delta_\theta^\pm],
\end{equation}
where $\Delta_\theta^\pm=1\mp 2Am+a^2 A^2$. The deficit angles are produced by the presence of the cosmic strings along the axes, which provide tensions $\mu_\pm=\varpi_\pm/(8\pi)$. The acceleration of the black hole can be viewed as  a mismatch of conical deficits between the two poles of the black hole, as $\mu_- -\mu_+=CAm>0$ \cite{Appels:2016uha,Anabalon:2018ydc,Anabalon:2018qfv,Zhang:2019vpf}. As a result, the phenomena that the latitude of the photon's circular orbit will be closer to the south pole and also the observer's critical inclination angles tend to the south pole can be attributed to the mismatch of the cosmic tension from the south pole and the north pole.

\section*{Acknowledgements}
We thank Li Jing for helpful discussions in the initial stage of this work. Jie Jiang  is supported by the National Natural Science Foundation of China (Grants No. 11775022 and 11873044). Ming Zhang acknowledges the start-up funding of Jiangxi Normal University.

\end{document}